\documentclass[aip,jap,print,graphicx]{revtex4-1}

\usepackage{graphicx}
\usepackage{dcolumn}
\usepackage{bm}

\usepackage{amsmath}
\usepackage{amssymb}

\begin{document}


\title{Studying of the interlayer interaction in magnetic multilayers (FM/I/FM) measuring the FMR peak asymmetry} 



\author{S.N.~Vdovichev}
\affiliation{Institute for Physics of Microstructures RAS, Nizhny Novgorod, Russia}

\author{N.S.~Gusev}
\affiliation{Institute for Physics of Microstructures RAS, Nizhny Novgorod, Russia}

\author{S.A.~Gusev}
\affiliation{Institute for Physics of Microstructures RAS, Nizhny Novgorod, Russia}

\author{L.I. Budarin}
\affiliation{Lobachevsky State University of Nizhny Novgorod, Nizhny Novgorod, Russia}

\author{D.A. Tatarskiy}
\affiliation{Institute for Physics of Microstructures RAS, Nizhny Novgorod, Russia}
\affiliation{Lobachevsky State University of Nizhny Novgorod, Nizhny Novgorod, Russia}

\author{O.L.~Ermolaeva}
\affiliation{Institute for Physics of Microstructures RAS, Nizhny Novgorod, Russia}

\author{V.V.~Rogov}
\affiliation{Institute for Physics of Microstructures RAS, Nizhny Novgorod, Russia}

\author{O.G.~Udalov}
\email[]{oleg.udalov@csun.edu}
\affiliation{Department of Physics and Astronomy, California State University Northridge, Northridge, California 91330, USA}
\affiliation{Institute for Physics of Microstructures RAS, Nizhny Novgorod, Russia}

\author{I.S.~Beloborodov}
\affiliation{Department of Physics and Astronomy, California State University Northridge, Northridge, California 91330, USA}

\author{E.S.~Demidov}
\affiliation{Lobachevsky State University of Nizhny Novgorod, Nizhny Novgorod, Russia}

\author{A.A.~Fraerman}
\affiliation{Institute for Physics of Microstructures RAS, Nizhny Novgorod, Russia}


\date{\today}

\begin{abstract}
We experimentally study the interlayer interaction in a magnetic multilayer system ferromagnet/insulator/ferromagnet with different spacer thickness. We demonstrate that the sign and the magnitude of the interaction can be deduced from the FMR peak shape rather than from the FMR peak shift. The proposed technique allows studying the interlayer interaction using a single sample (without a reference sample for comparison).
\end{abstract}


\maketitle 

\section{Introduction}\label{Sec:Intro}
A magnetic tunnel junction (MTJ) is in the focus of spintronics promising several interesting applications~[\onlinecite{Hillebrands2014,Fert2010,Ohno2007,Park2006,Edelstein2006}]. The MTJ consists of two ferromagnetic (FM) layers separated by an insulating (I) spacer. An interaction between the layers in the MTJ defines a system ground state~[\onlinecite{Tsymbal2006,Jonge1997,Schuhl2002,Lesnik2007,Yi2010,Baberschke2003,Nozieres2004,Parkin2000,Lin1999,Fraerman2018,Schneider2001,Powell2006}]. It influences a susceptibility of magnetic field sensors based on MTJ systems. The interlayer interaction also plays a crucial role in magnetization switching processes related to information writing in MTJ based memory.  

To investigate the interlayer interaction people often use the ferromagnetic resonance (FMR). Usually, the interaction between magnetic layers is studied by measuring of the FMR peaks shift. This is quite difficult since the FMR peak shift (in the case of tunnel junction) is small comparing to the FMR peak width. Moreover, a reference sample is always needed to define the shift of the peaks.

Recently, another method for defining the interlayer coupling sign and magnitude was theoretically proposed~[\onlinecite{Demidov2018}]. This method is based on analysing of FMR peaks shape rather than shift. In particular, according to Ref.~[\onlinecite{Demidov2018}] the interlayer interaction leads to the appearance of the FMR peak asymmetry. Such an asymmetry occurs only when FMR peaks corresponding to two magnetic layers of MTJ overlap.  

Advantage of this method is related to the fact that there is no need to use a reference sample or several samples with different thickness of the insulating spacer. A single sample can be studied and the interlayer interaction can be obtained.

In the present paper we study a series of MTJs. At first we use ``traditional'' methods for studying of the interlayer coupling between magnetic layers such as magneto-optical Kerr effect (MOKE) and FMR method based on the shift of the FMR peaks. This allows us to confirm existence of the interlayer interaction and estimate its sign and magnitude. After that we perform specific measurements of FMR peaks shape and observe the peak asymmetry. Using these measurements we show that interlayer coupling can be deduced from the FMR peak shape.

The paper is organized as follows. Sec.~\ref{Sec:Exp} describes all experimental procedures. Theoretical background and modelling procedures are described in Sec.~\ref{Sec:Disc}. Discussion of experimental results are given in the second part of Sec.~\ref{Sec:Disc}.

\begin{figure}
	\includegraphics[width=0.5\columnwidth]{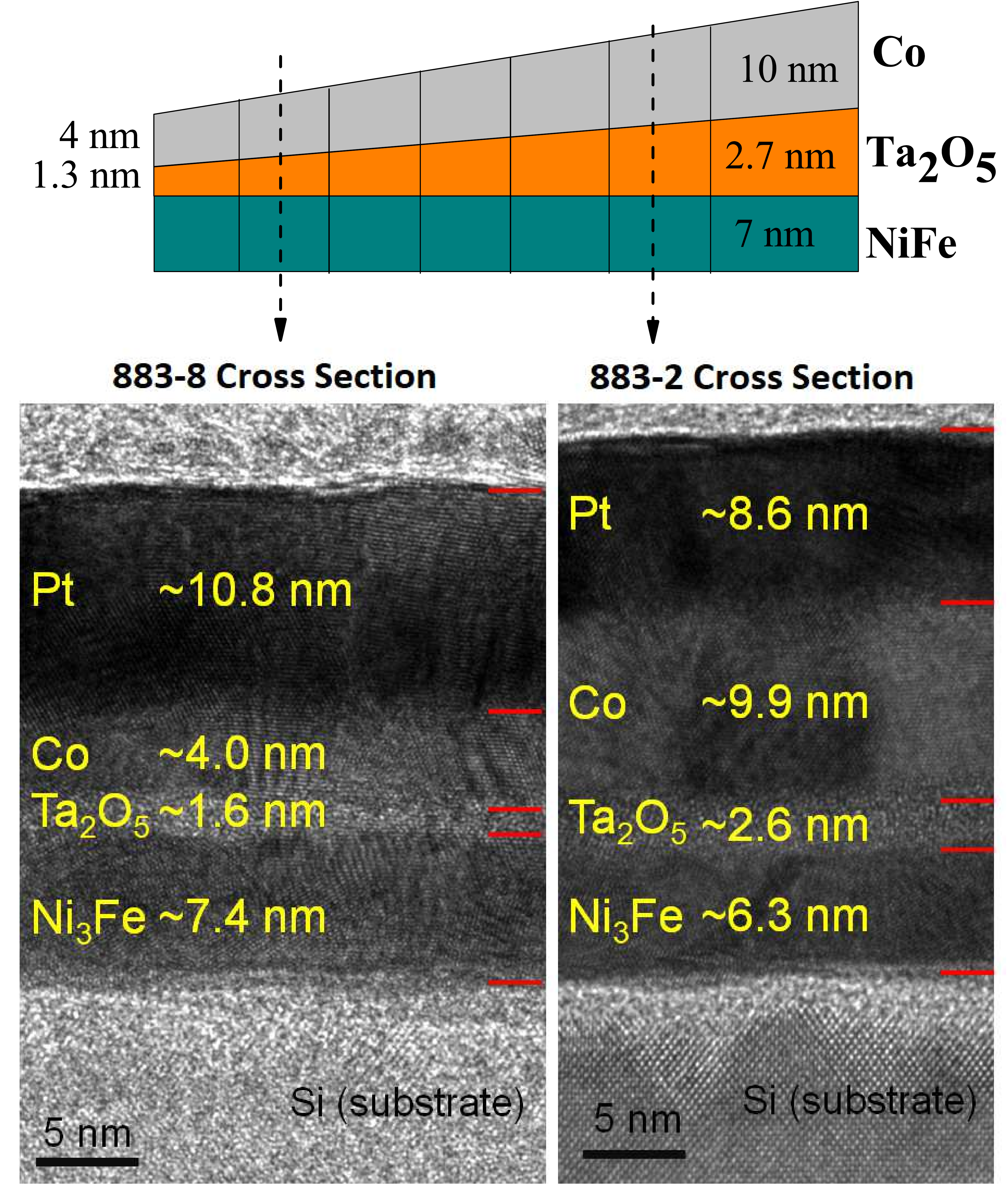}
	\caption{Upper panel: cartoon picture of samples. Lower panels: TEM images of two samples with different Ta$_2$O$_5$ thickness.\label{Fig:Samples}}%
\end{figure}

\section{Experimental and modeling procedures}\label{Sec:Exp}

\subsection{Fabrication technique. Samples description}

Magnetic multilayer structure Ni$_{80}$Fe$_{20}$(7 nm)/Ta$_2$O$_5$(1.4 - 2.7 nm)/Co(4 - 10 nm)/Pt(10 nm) was deposited at room temperature on silicon substrates using magnetron sputtering system AJA ATC2200. The base pressure in the main chamber was $\sim5\cdot10^{-8}$ Torr, the working pressure was 2 mTorr. The substrate was cleaned by Ar  plasma before deposition of the structure in loadlock chamber. The metallic layers were fabricated in Ar atmosphere. The substrate was rotated (30 rpm) during the deposition of the NiFe layer. The thickness of NiFe layer is about 7 nm. The Ta$_2$O$_5$ layer was deposited in a mixed atmosphere of Ar and O$_2$ using a metallic Ta target. The chamber was pumped up to a ground pressure before sputtering of the Co layer. Sputtering of the Ta$_2$O$_5$ and Co layers was performed without rotation. This allows to fabricate the wedge Ta$_2$O$_5$ layer (see Fig.~\ref{Fig:Samples} upper panel). The thickness of Co layer varied also. Sputtering without rotation induces uniaxial in-plane anisotropy in the Co layer. The wedge sample were cut into several pieces with different thickness of the Ta$_2$O$_5$ layer from 1.4 to 2.7 nm. 

Transition electron microscopy (TEM) was used to check thickness of the layers. TEM images for samples with thick and thin insulating spacer layer are shown in Fig.~\ref{Fig:Samples} (lower panels). NiFe thickness is about 7 nm in both samples. Co thickness decreases from 10 nm to 4 nm with decreasing of the spacer thickness. The insulating spacer thickness changes from 2.6 nm to 1.6 nm. Important to mention that there are no pinholes in the images and the spacer is more or less uniform. There is no evident correlation between Co/Ta$_2$O$_5$ and  NiFe/Ta$_2$O$_5$ interfaces. To check this we study images with longer length. This is important because the ``orange-peel'' effect appears only for films with correlated roughness.

\subsection{Measurement techniques}

The cross sections for high resolution transmission microscopy (HRTEM) were prepared as lamellas using Ga$^+$ 30 keV ions in the cross-beam SEM-FIB workstation Zeiss AURIGA (Interdisciplinary resource center for nanotechnology, Saint Petersburg, Russia). High energy ions created a very thick damaged amorphous layer on the lamella sides. Therefore, the lamellas were additionally polished by low-energy ions Ar$^+$ 0.5 keV to reduce damaged amorphous layer. HRTEM measurements were performed with a LIBRA 200 MC Shottky Field emission gun instrument operating at 200 kV. The scale calibration was done using Si (111) substrate, visible on HRTEM micrographs. The micrographs was averaged over horizontal direction to extract quantitative information about layers thickness.

The morphology of the films was studied by the atomic-force microscopy (AFM, “Solver-HV,” NT-MDT). 

A magneto-optical Kerr effect (MOKE) for the samples was measured with a home-built system. We used meridional geometry. A He-Ne (wavelength 632 nm, 5 mW power) laser with linear polarization was used as a light source. The samples were mounted inside a gap of an electromagnet which allowed magnetic fields of up to 3 KOe to be applied in the plane of the sample. During the measurement, data were taken as a function of magnetic field to generate a hysteresis loop. We measured a full hysteresis loop at first. Our samples consist of two magnetic films with essentially different coercivity. This allows us to study a so-called minor loop of the magnetically soft NiFe layer. To get the minor loop we started measurements at high negative field. We increased the field until we switched the NiFe layer. After that we decreased the field back to high negative value without switching the hard Co layer. 

\begin{figure}
	\includegraphics[width=0.5\columnwidth]{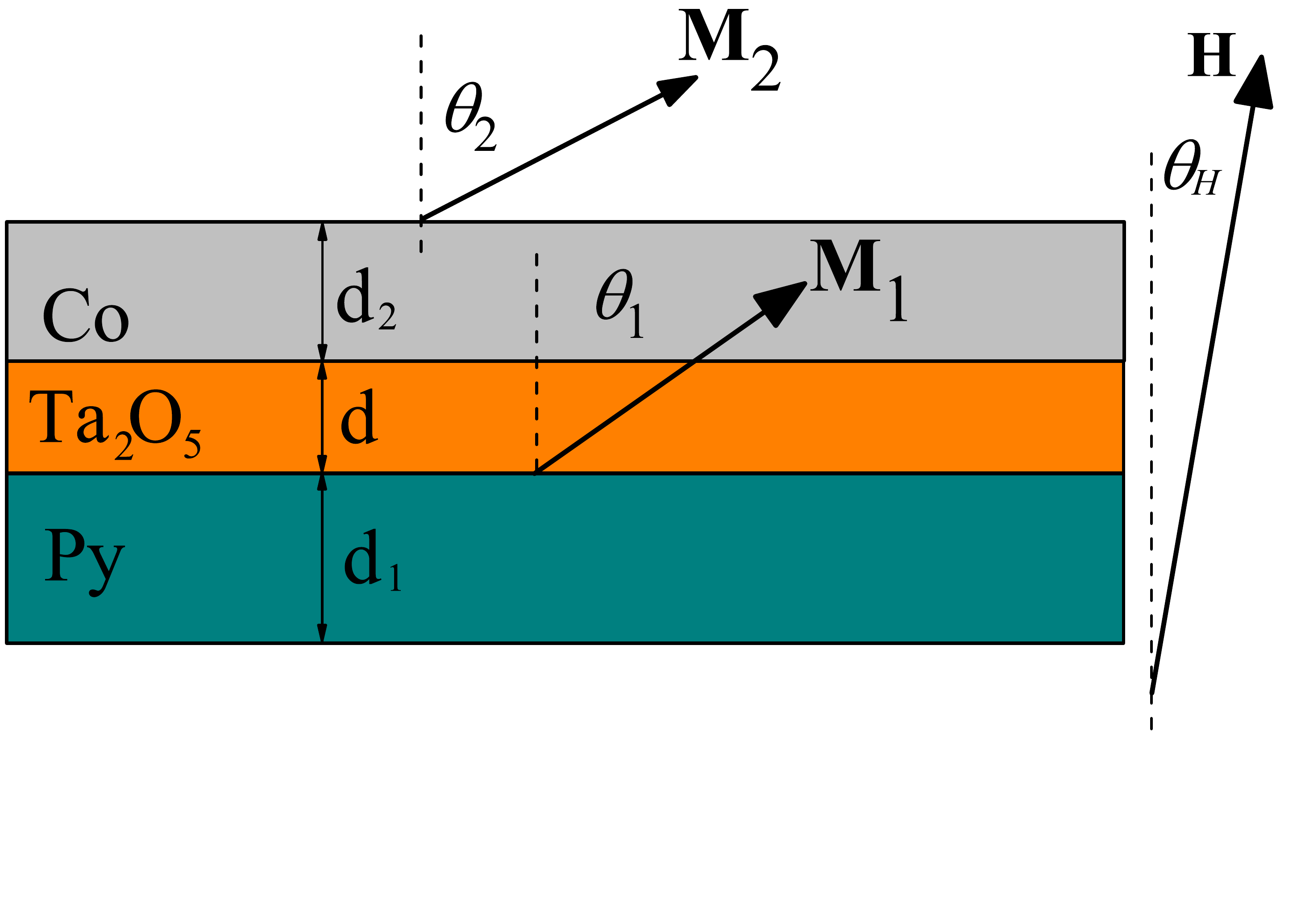}
	\caption{Cartoon picture of a single sample. Magnetizations $\mathbf M_{1,2}$ make agles $\theta_{1,2}$ with the MTJ plane. The external magnetic field $\mathbf H$ is inclined with respect to the sample normal by an angle $\theta_H$.\label{Fig:OPexpl}}%
\end{figure}

The FMR measurements on fabricated MTJs were performed at room temperature with Bruker EMX Plus-10/12 spectrometer equipped by dc magnet with field $H$ up to 1.5 T. The polarized microwave magnetic field $\mathbf h$ with frequency 9.8 GHz (TE$_{011}$ mode of the cylindrical resonant cavity) was perpendicular to the field $\mathbf H$. The samples were driven through the resonance by the magnitude of magnetic field $H$ sweeping. Two types of measurement were used. In the first experiment we applied external field along the MTJ plane. We studied a field dependence of the absorbed power $W(H)$. In the second experiment, we measured $W(H)$ when the magnetic field is inclined with respect to the sample plane. We introduce here the angle $\theta_H$ between the applied field $\mathbf H$ and the sample normal (see Fig. \ref{Fig:OPexpl}). This angle is chosen as explained below. Our MTJs consist of two different magnetic films. Therefore, there are two peaks in $W(H)$. These peaks appear at resonant fields $H_\mathrm{res}^{(1,2)}$. Magnitudes of the resonant fields depend on the inclination angle $\theta_H$. At a certain angle $\alpha_\mathrm{cr}$ these fields are equal $H_\mathrm{res}^{(1)}=H_\mathrm{res}^{(2)}$. In our study $\theta_H^\mathrm{cr}$ is about 5 deg. We study FMR spectrum thoroughly close to the critical angle $\theta_H^\mathrm{cr}$.

\subsection{FMR spectrum modelling procedure}
We use a well known numerical algorithm to solve the LLG equations for magnetic films~[\onlinecite{Baberschke2003,Ozdemir2010,Furdyna2006}]. The system energy is given by

\begin{equation}\label{Eq:SysEn}
E=E_\mathrm Z+E_\mathrm D+E_{\mathrm A}+E_{\mathrm{int}},
\end{equation}
where the Zeeman energy is 
\begin{equation}\label{Eq:EnZee}
E_\mathrm Z=-\sum_{i=1,2}d_i(\mathbf M_i\mathbf H),
\end{equation}
magneto-dipole shape anisotropy is
\begin{equation}\label{Eq:EnShape}
E_\mathrm D=\sum_{i=1,2}2\pi d_i M_i^2\cos^2(\theta_i),
\end{equation}
uniaxial anisotropy is
\begin{equation}\label{Eq:EnAn}
E_\mathrm A=\sum_{i=1,2}d_i K^{(2)}_i\cos^2(\theta_i)+\sum_{i=1,2}d_i K^{(4)}_i\cos^4(\theta_i).
\end{equation}
We consider here the case of isotropic exchange coupling which is given by
\begin{equation}\label{Eq:EnInt}
E_\mathrm{int}=-\tilde J(\mathbf M_1 \mathbf M_2).
\end{equation}
External magnetic field $\mathbf H$ is inclined by an angle $\theta_\mathrm H$ with respect to the sample normal. $K$ is the anisotropy constant, $J$ is the coupling constant (we will discuss different kind of magnetic interaction in MTJ in Sec.~\ref{Sec:Disc}). Equilibrium angles of magnetizations (at $\mathbf h=0$) are defined by minimization of the system energy Eq.~(\ref{Eq:SysEn}). 

Using experimental dependencies of resonance field $H_\mathrm{res}$ for NiFe and Co layers on the field angle $\theta_H$ we define the parameters of magnetic films. In particular the best fit is obtained when saturation magnetization of the films are $M_{Co}=1420$ Gs, $M_{NiFe}=500$ Gs, anisotropy constants $K^{(2)}_1=4.95\cdot 10^6$ Gs$\cdot$Oe, $K^{(2)}_2=-6.5\cdot 10^5$ Gs$\cdot$Oe, $K^{(4)}_1=1.65\cdot 10^6$ Gs$\cdot$Oe, $K^{(4)}_2=-1.4\cdot 10^5$ Gs$\cdot$Oe, damping parameters $\alpha_1=0.046$ and $\alpha_2=0.01$, g-factors $g_1=2$, $g_2=2.1$

\subsection{Defining the interlayer interaction sign from the FMR peak shape}\label{Sec:TheoryShape}
According to Ref.~[\onlinecite{Demidov2018}] the FMR peak shape contains the information on the interlayer interaction. One can define the interaction sign when two FMR peaks corresponding to two magnetic layers overlap. Changing the angle of external magnetic field $\theta_H$ one can always find the field direction at which the resonant fields of both peaks are the same. In this case the Fano resonance appears leading to skewing of the joint FMR peak corresponding to the layer with smaller dissipation (in our case this is NiFe layer). If this narrow peak has higher slope at the lower field (at the left side) then the interaction is of FM type. If the slope is higher at the right part of the peak then there is an AFM interaction between the layers. Modelling the peak shape one can even estimate the magnitude of the interlayer interaction.

\section{Discussion and analysis}\label{Sec:Disc}

\subsection{AFM measurements}
Using AFM we study the surface roughness of the upper Co layer in the fabricated samples. Since layers thickness in our samples is quite small one can safely suggest that roughness of all interfaces in the sample is the same. We get the root-mean square roughness (roughness height) of order of $\sigma=0.3$ nm and the lateral correlation length of roughness of order of $\lambda=30$ nm. 

\begin{figure}
	\includegraphics[width=1\columnwidth]{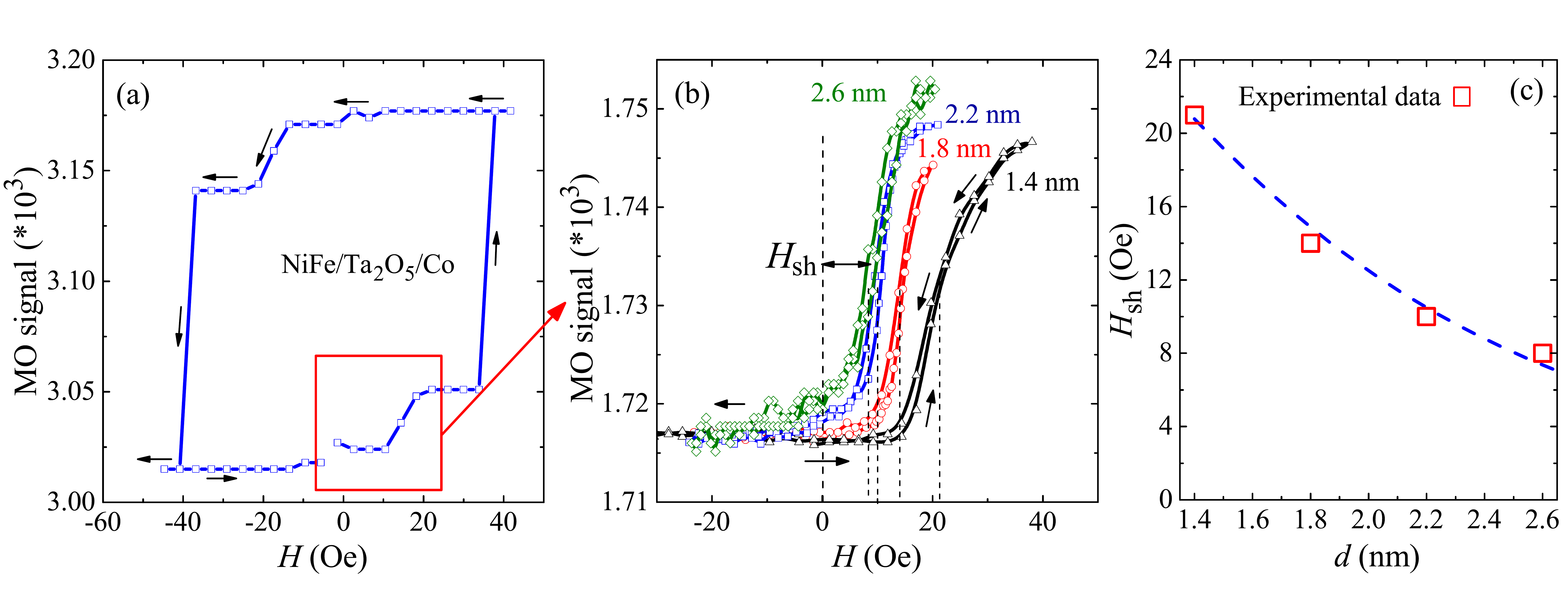}
	\caption{(a) Typical hysteresis loop of NiFe/Ta$_2$O$_5$/Co multilayer. Arrows show the bypass direction. (b) Minor hysteresis loops for samples with different thickness of the insulating spacer $d=2.6$, $2.2$, $1.8$ and $1.4$ nm. The minor loops are shifted with respect to zero by the field $H_\mathrm{sh}$. (c) Dependence of the minor loops shift $H_\mathrm{sh}$ on the oxide layer thickness $d$. Open squares show the experimental data. Blue dashed line is the guide for eyes showing expeonentially deacaying function. \label{Fig:MO_Sample1}}%
\end{figure}

\subsection{Magneto-optical measurements. Thickness dependence of the interlayer interaction}

Figure~\ref{Fig:MO_Sample1} shows the results of the MOKE studies for samples with different insulator spacer thickness. Left panel shows the typical full hysteresis loop. All samples show such a loop. At high field ($H>50$ Oe) both Co and NiFe layers are magnetized along the field and the system is in the FM state. This state preserves until one reaches the negative field of $H=-20$ Oe, at which the NiFe layer switches. In the field region $-40<H<-20$ Oe the system state is AFM. Switching of the Co layer happens at $H=-40$ Oe transforming the system into FM state again.

To study the interaction between magnetic layers we measure minor hysteresis loops which are shown in the central panel. The width of the minor loops is of order of 5 Oe which corresponds to the NiFe film coercive field. The minor loops for all thicknesses are shifted toward the switching field of the Co layer. This means that there is a FM interaction between the Co and NiFe layers [\onlinecite{Jonge1997}]. The shift $H_\mathrm{sh}$ decreases with increasing the insulator thickness $d$ (see the right panel of Fig.~\ref{Fig:MO_Sample1}). So, the MOKE measurements show that there is an interaction between magnetic layers decreasing with increasing of the insulator spacer thickness $d$.

\subsection{Ferromagnetic resonance}
Ferromagnetic resonance is a well known technique for studying of the interlayer coupling~[\onlinecite{Baberschke2003,Passamani2006,Lesnik2006,Lesnik2007,PhysRevApplied.8.044006}]. However, in most cases the interaction is studied for the case of the in-plane magnetization. Usually, a magnetic bilayer system shows two FMR peaks. Mutual shift of these peaks provides the information on the interlayer coupling. Mutual shift can be defined only if one has some reference sample without interaction or as in our case several samples with different spacer thickness. 

\begin{figure}
	\includegraphics[width=0.5\columnwidth]{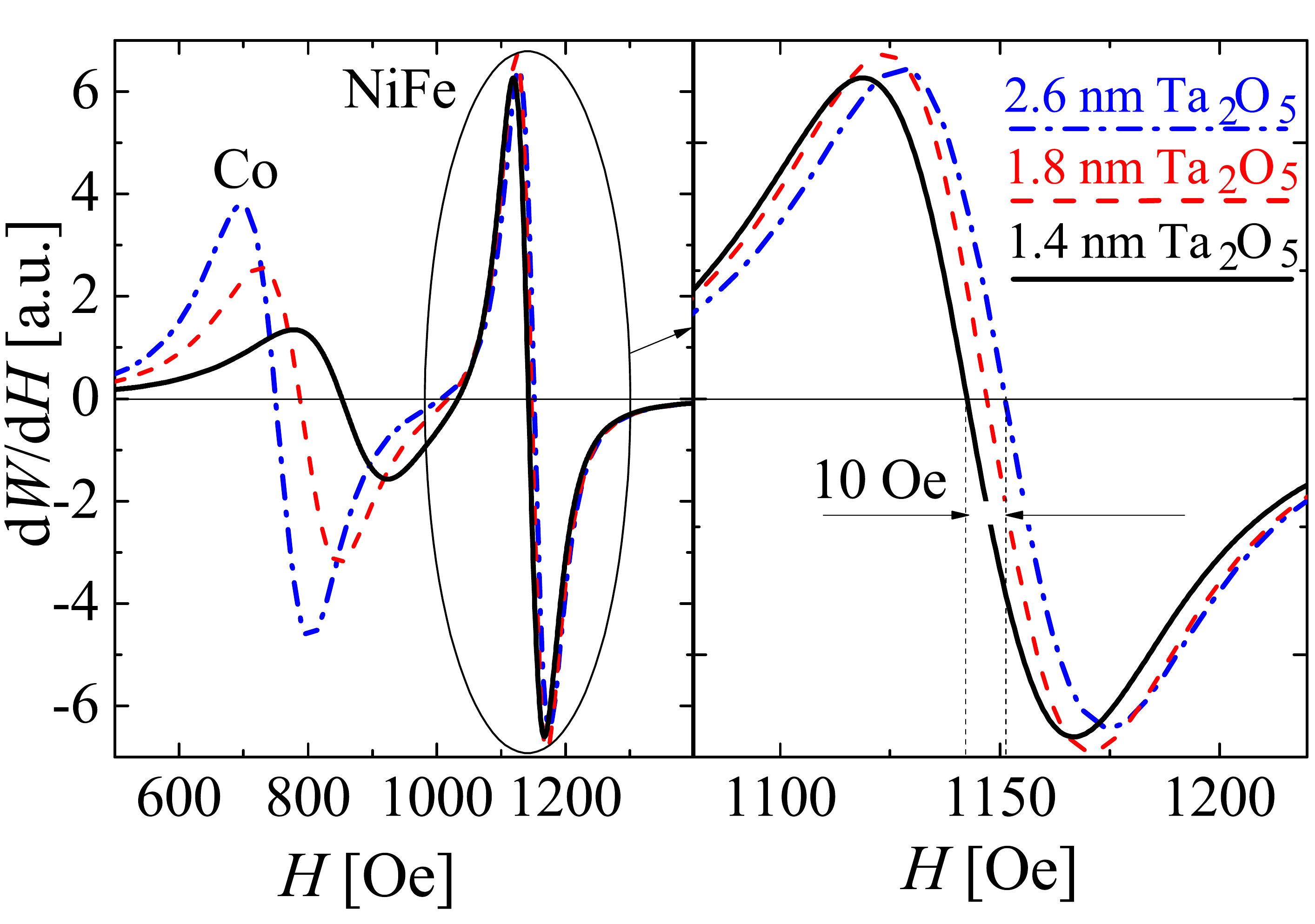}
	\caption{FMR signal  $dW/dH$ as a function of the external field $H$ for samples with different thickness of the insulating spacer $d$. Left panel: the full spectrum. Right panel: the NiFe peak. Black solid line is for sample with Ta$_2$O$_5$ thickness of order of $d=1.4$ nm. Red dashed line is for $d=1.8$ nm. Blue dash-dot-dotted line corresponds to $d=2.6$ nm. \label{Fig:FMR883inplane}}
\end{figure}

We perform such ``conventional'' in-plane measurements to further confirm existence of the interlayer interaction in our system. In the case of the in-plane measurements we use traditional way to obtain the sign of interaction.
Figure~\ref{Fig:FMR883inplane} shows FMR signal $dW/dH$ (the field derivative of the absorbed power $W(H)$) for the several samples with different Ta$_2$O$_5$ thickness. The quantity $dW/dH$ easily allows  to find the FMR peaks position (the resonant field $H_\mathrm{res}^{(1,2)}$). They are defined as points where $dW/dH=0$. Left panel shows the FMR spectrum in a wide range of the external field. Two ``peaks'' (instead of a peak one can see a kink since we plot the derivative $dW/dH$) are visible. The low field ($H_\mathrm{res}^{(1)}\approx 800$ Oe) peak corresponds mainly to the Co layer, while the peak at $H_\mathrm{res}^{(2)}\approx 1150$ Oe is due to the NiFe film. The FMR spectrum for samples with different thickness of the Ta$_2$O$_5$ layer $d$ are shown by different lines in Fig.~\ref{Fig:FMR883inplane}. One can see that decreasing of the spacer thickness leads to shifting of the peaks closer to each other. Note that due to specific fabrication technique the samples with thinner Ta$_2$O$_5$ layer have thinner Co layer. According to our simulations, the shift of the Co peak is mostly due to the reduction of the Co layer thickness. At the same time the thickness of the NiFe layer is the same for all samples. The shift of the NiFe peak toward the Co one means that there is a FM interaction growing with decreasing of the insulator spacer thickness $d$. The shift is of order of 10 Oe. This is in agreement with the data of MOKE measurements (see Fig.~\ref{Fig:MO_Sample1}). 

\subsubsection{Defining the interaction sign and magnitude from the FRM peak shape}

\begin{figure}
	\includegraphics[width=0.5\columnwidth]{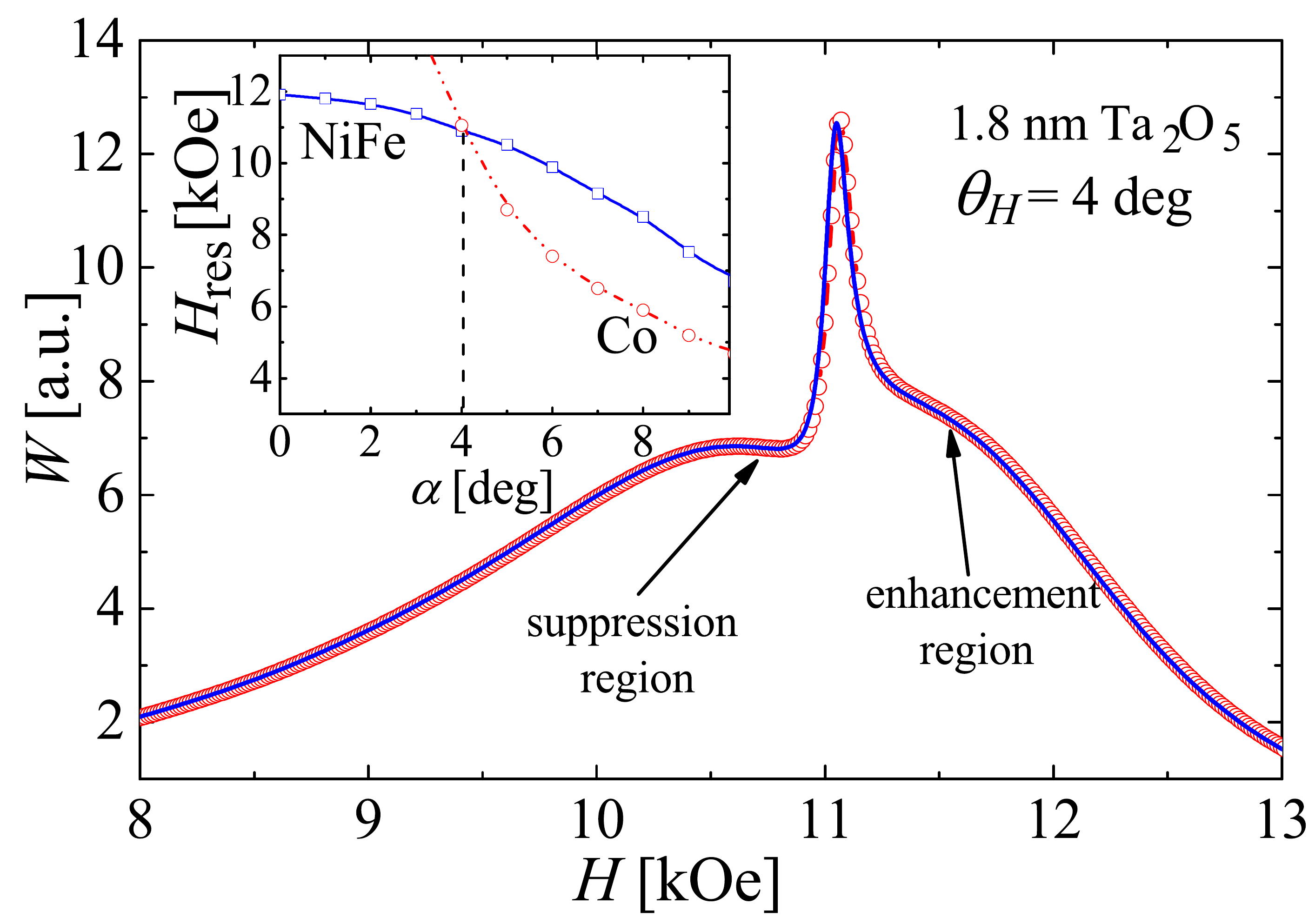}
	\caption{FMR signal $W$ (the absorbed power) as a function of external field magnitude $H$. The external magnetic field makes angle $\theta_H=4$ deg with the sample normal. Red open sqares are the experimental data for sample with Ta$_2$O$_5$ thickness of order of $d=1.8$ nm. Blue solid line shows numerical modeing of FMR signal. Inset shows NiFe (blue solid) and Co (red dashed)  resonance fields $H_\mathrm{res}$ as a function of the angle $\theta_{H}$ for the sample with $d=1.8$ nm.\label{Fig:FMRoutofplane883}}%
\end{figure}

As we discussed in Sec.~\ref{Sec:TheoryShape} there was recently proposed another method for studying of the interlayer interaction. According to this method we measure angular dependence of the NiFe and Co peaks positions $H_\mathrm{res}^{(1,2)}$ as a function of the angle $\theta_H$. Such a dependence for the sample with $d=1.8$ nm is plotted in the inset in Fig.~\ref{Fig:FMRoutofplane883}. Similar dependencies are obtained for all other samples. The resonant fields dependencies intersect at a critical angle $\theta^\mathrm{cr}_H=4$ deg. The main plot in Fig.~\ref{Fig:FMRoutofplane883} represents the FMR spectrum for the sample with 1.8 nm spacer at a critical angle $\theta^\mathrm{cr}_H=4$ deg.  Note that the critical angle for all sample is rather small meaning that we apply external field almost perpendicular to the MTJ plane. 

In contrast to the previous Fig.~\ref{Fig:FMR883inplane}, here we plot the absorbed power $W(H)$ itself. This allows to analyse a peak shape. Red circles in the main plot of Fig.~\ref{Fig:FMRoutofplane883} demonstrate the experimental data. One can see that the narrow (NiFe) peak is asymmetric evidencing the interlayer interaction. The higher slope is at the left side of the peak. This means that the observed interaction is of FM type. We perform numerical modelling to fit the experimental data. The results of numerical simulations are shown with a blue solid line. One can see a good agreement between experimental and theoretical curves. Modelling shows that interaction strength is of the order of  $J=0.01$ erg/cm$^2$.  This interaction strength gives the effective field of 30 Oe acting on the NiFe film. This is of order of the shift of the minor hysteresis loop in our MOKE measurements. However, it exceeds MOKE shift. The reason for this discrepancy requires further investigations.

\begin{figure}
	\includegraphics[width=0.5\columnwidth]{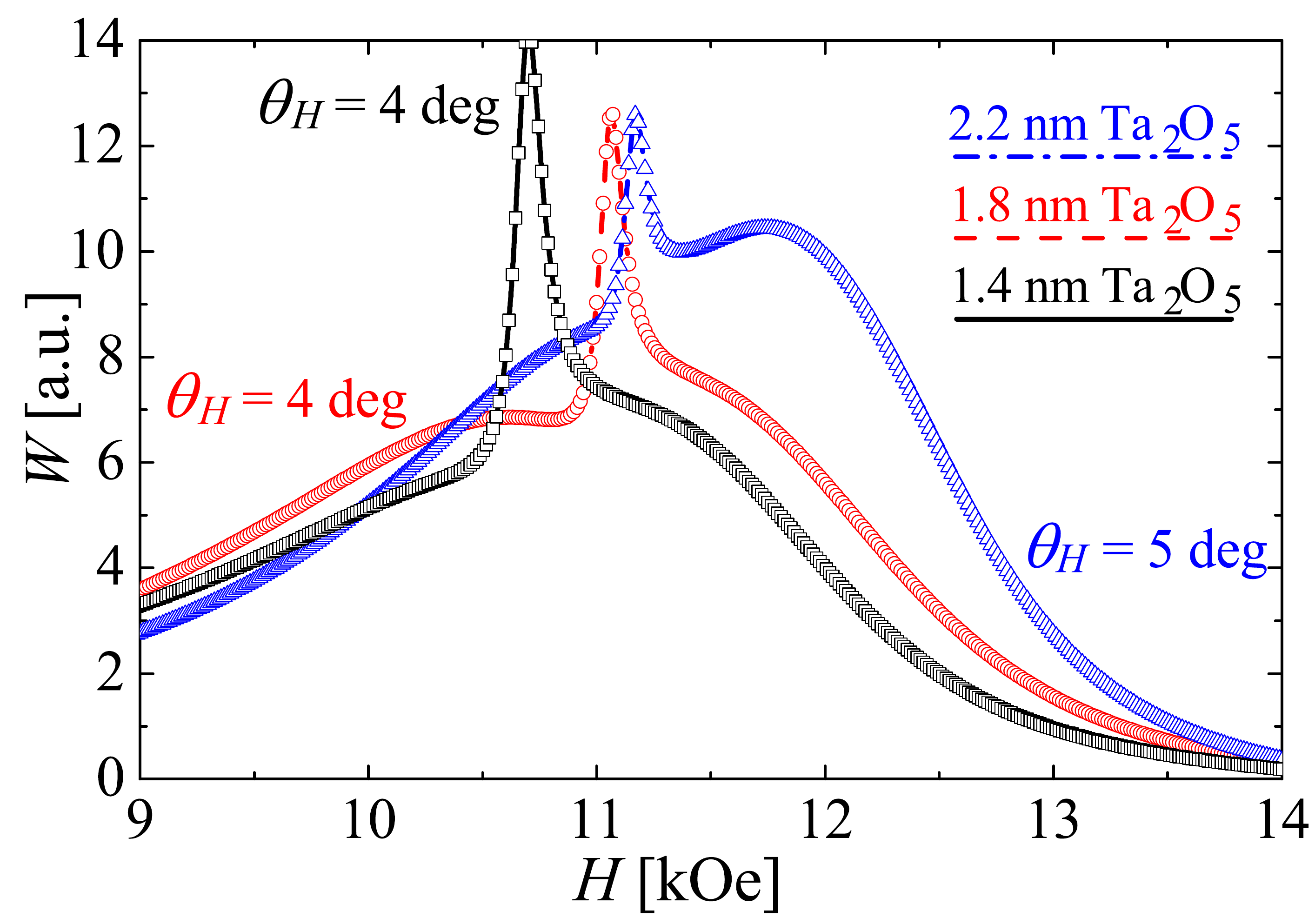}
	\caption{FMR signal $W$ (the absorbed power) as a function of external field magnitude $H$. Black open squares are for the sample with  Ta$_2$O$_5$ thickness of order of $d=1.4$ nm and $\theta_H=4$ deg. Red open circles are for the sample with  the thickness $d=1.8$ nm and $\theta_H=4$ deg. Blue open triangles are for the sample with  the thickness $d=2.2.$ nm and $\theta_H=5$ deg.\label{Fig:FMRoop2}}%
\end{figure}

Other samples demonstrate similar dependencies of the FMR peaks. They are shown in Fig.~\ref{Fig:FMRoop2}. One can see that the mentioned asymmetry can be easily seen in samples with $d=1.4$ and $d=1.8$ nm. The asymmetry is not seen in the sample with $d=2.2$ nm. This agrees with decreasing of the  interaction with increasing of the insulating spacer. 

\subsection{Discussion}\label{Sec:Disc}

In previous sections we did not discuss the origin of the interlayer coupling. There are a few possible types of the interaction: 1) the interlayer exchange coupling~[\onlinecite{Tsymbal2006,Jonge1997,Schuhl2002,Lesnik2007,Yi2010,Baberschke2003}] ; 2) the magneto-dipole ``orange-peel'' (OP) effect~[\onlinecite{Nozieres2004,Parkin2000,Lin1999,Fraerman2018,Schneider2001,Powell2006}]; 3) and the coupling due to pin-holes. For the multilayer structure of good qulity one can neglect the pin-holes. Distinguishing of the exchange coupling and the ``orange-peel'' effect is a challenging task. Both of them decays exponentially and can be of the same order. Therefore, thickness dependence of the interlayer coupling can not be used for distinguishing of these interactions on a qualitative level. Our estimates based on the AFM data favours the exchange coupling in our samples (roughness is quite low and smooth, so the OP interaction should be rather small). 

Here we would like to mention that there is a qualitative difference between the exchange and OP interactions. The exchange coupling is isotropic, while the OP effect is anisotropic~[\onlinecite{Fraerman2018}]. Generally, this peculiarity can be used for distinguishing between these interactions using the method based on the FMR peaks shape analysis. Anisotropy of the OP effect should lead to the dependence of the coupling on the angle $\theta_H$. Isotropic exchange coupling should be independent of $\theta_H$. Thus, performing more subtle angular ($W(H,\theta_H)$) measurements one can provide an information on the interaction type. This question requires further investigations.

Finally, we would like to mention that the difference in the interaction strength obtained by MOKE and FMR methods can be related to the anisotropy of the OP effect. MOKE measurements were done in the in-plane geometry, while the FMR peak shape were studied in the out-of-plane geometry.  Thus, the contribution to the total interaction from the OP effect can be different.

\section{Conclusion}

We experimentally studied the interlayer interaction in a magnetic multilayer system with two ferromagnetic layers separated by an insulating spacer. Several samples with different thickness of the insulating spacer were investigated. We proposed the method for defining the sign and the shape of the interaction based on analyses of FMR peaks shape rather than peaks shift. This method is based on studying of the FMR spectra of the system at different angles of an external field. At a certain angle FMR peaks of both magnetic layers overlap. At that the FMR peak becomes asymmetric. If the peak has higher slope at the left side there is a FM interaction. Oppositely, if the peak has higher slope at the right side the interaction is AFM. Numerical modelling of FMR signal allows to define the magnitude of interaction. This method allowed us to find the interlayer coupling in NiFe/Ta$_2$O$_5$/Co system.

\section{Acknowledgments}

This research was supported was supported by the Russian Science Foundation (Grant  16-12-10340). O.~U. and I.~B. were supported by NSF under Cooperative Agreement Award EEC-1160504.

\bibliography{ExExp}

\begin{thebibliography}{23}%
\makeatletter
\providecommand \@ifxundefined [1]{%
 \@ifx{#1\undefined}
}%
\providecommand \@ifnum [1]{%
 \ifnum #1\expandafter \@firstoftwo
 \else \expandafter \@secondoftwo
 \fi
}%
\providecommand \@ifx [1]{%
 \ifx #1\expandafter \@firstoftwo
 \else \expandafter \@secondoftwo
 \fi
}%
\providecommand \natexlab [1]{#1}%
\providecommand \enquote  [1]{``#1''}%
\providecommand \bibnamefont  [1]{#1}%
\providecommand \bibfnamefont [1]{#1}%
\providecommand \citenamefont [1]{#1}%
\providecommand \href@noop [0]{\@secondoftwo}%
\providecommand \href [0]{\begingroup \@sanitize@url \@href}%
\providecommand \@href[1]{\@@startlink{#1}\@@href}%
\providecommand \@@href[1]{\endgroup#1\@@endlink}%
\providecommand \@sanitize@url [0]{\catcode `\\12\catcode `\$12\catcode
  `\&12\catcode `\#12\catcode `\^12\catcode `\_12\catcode `\%12\relax}%
\providecommand \@@startlink[1]{}%
\providecommand \@@endlink[0]{}%
\providecommand \url  [0]{\begingroup\@sanitize@url \@url }%
\providecommand \@url [1]{\endgroup\@href {#1}{\urlprefix }}%
\providecommand \urlprefix  [0]{URL }%
\providecommand \Eprint [0]{\href }%
\providecommand \doibase [0]{http://dx.doi.org/}%
\providecommand \selectlanguage [0]{\@gobble}%
\providecommand \bibinfo  [0]{\@secondoftwo}%
\providecommand \bibfield  [0]{\@secondoftwo}%
\providecommand \translation [1]{[#1]}%
\providecommand \BibitemOpen [0]{}%
\providecommand \bibitemStop [0]{}%
\providecommand \bibitemNoStop [0]{.\EOS\space}%
\providecommand \EOS [0]{\spacefactor3000\relax}%
\providecommand \BibitemShut  [1]{\csname bibitem#1\endcsname}%
\let\auto@bib@innerbib\@empty
\bibitem [{\citenamefont {Stamps}\ \emph {et~al.}(2014)\citenamefont {Stamps},
  \citenamefont {Breitkreutz}, \citenamefont {Akerman}, \citenamefont {Chumak},
  \citenamefont {Otani}, \citenamefont {Bauer}, \citenamefont {Thiele},
  \citenamefont {Bowen}, \citenamefont {Majetich}, \citenamefont {Klaui},
  \citenamefont {Prejbeanu}, \citenamefont {Dieny}, \citenamefont {Dempsey},\
  and\ \citenamefont {Hillebrands}}]{Hillebrands2014}%
  \BibitemOpen
  \bibfield  {author} {\bibinfo {author} {\bibfnamefont {R.~L.}\ \bibnamefont
  {Stamps}}, \bibinfo {author} {\bibfnamefont {S.}~\bibnamefont {Breitkreutz}},
  \bibinfo {author} {\bibfnamefont {J.}~\bibnamefont {Akerman}}, \bibinfo
  {author} {\bibfnamefont {A.~V.}\ \bibnamefont {Chumak}}, \bibinfo {author}
  {\bibfnamefont {Y.}~\bibnamefont {Otani}}, \bibinfo {author} {\bibfnamefont
  {G.~E.~W.}\ \bibnamefont {Bauer}}, \bibinfo {author} {\bibfnamefont {J.-U.}\
  \bibnamefont {Thiele}}, \bibinfo {author} {\bibfnamefont {M.}~\bibnamefont
  {Bowen}}, \bibinfo {author} {\bibfnamefont {S.~A.}\ \bibnamefont {Majetich}},
  \bibinfo {author} {\bibfnamefont {M.}~\bibnamefont {Klaui}}, \bibinfo
  {author} {\bibfnamefont {I.~L.}\ \bibnamefont {Prejbeanu}}, \bibinfo {author}
  {\bibfnamefont {B.}~\bibnamefont {Dieny}}, \bibinfo {author} {\bibfnamefont
  {N.~M.}\ \bibnamefont {Dempsey}}, \ and\ \bibinfo {author} {\bibfnamefont
  {B.}~\bibnamefont {Hillebrands}},\ }\href@noop {} {\bibfield  {journal}
  {\bibinfo  {journal} {J. Phys. D: Appl. Phys.}\ }\textbf {\bibinfo {volume}
  {47}},\ \bibinfo {pages} {333001} (\bibinfo {year} {2014})}\BibitemShut
  {NoStop}%
\bibitem [{\citenamefont {Dussaux}\ \emph {et~al.}(2010)\citenamefont
  {Dussaux}, \citenamefont {Georges}, \citenamefont {Grollier}, \citenamefont
  {Cros}, \citenamefont {Khvalkovskiy}, \citenamefont {Fukushima},
  \citenamefont {Konoto}, \citenamefont {Yakushiji}, \citenamefont {Yuasa},
  \citenamefont {Zvezdin}, \citenamefont {Ando},\ and\ \citenamefont
  {Fert}}]{Fert2010}%
  \BibitemOpen
  \bibfield  {author} {\bibinfo {author} {\bibfnamefont {A.}~\bibnamefont
  {Dussaux}}, \bibinfo {author} {\bibfnamefont {B.}~\bibnamefont {Georges}},
  \bibinfo {author} {\bibfnamefont {J.}~\bibnamefont {Grollier}}, \bibinfo
  {author} {\bibfnamefont {V.}~\bibnamefont {Cros}}, \bibinfo {author}
  {\bibfnamefont {A.}~\bibnamefont {Khvalkovskiy}}, \bibinfo {author}
  {\bibfnamefont {A.}~\bibnamefont {Fukushima}}, \bibinfo {author}
  {\bibfnamefont {M.}~\bibnamefont {Konoto}}, \bibinfo {author} {\bibfnamefont
  {H.~K.~K.}\ \bibnamefont {Yakushiji}}, \bibinfo {author} {\bibfnamefont
  {S.}~\bibnamefont {Yuasa}}, \bibinfo {author} {\bibfnamefont
  {K.}~\bibnamefont {Zvezdin}}, \bibinfo {author} {\bibfnamefont
  {K.}~\bibnamefont {Ando}}, \ and\ \bibinfo {author} {\bibfnamefont
  {A.}~\bibnamefont {Fert}},\ }\href@noop {} {\bibfield  {journal} {\bibinfo
  {journal} {Nature Communications}\ }\textbf {\bibinfo {volume} {1}},\
  \bibinfo {pages} {8} (\bibinfo {year} {2010})}\BibitemShut {NoStop}%
\bibitem [{\citenamefont {Ikeda}\ \emph {et~al.}(2007)\citenamefont {Ikeda},
  \citenamefont {Hayakawa}, \citenamefont {Lee}, \citenamefont {Matsukura},
  \citenamefont {Ohno}, \citenamefont {Hanyu},\ and\ \citenamefont
  {Ohno}}]{Ohno2007}%
  \BibitemOpen
  \bibfield  {author} {\bibinfo {author} {\bibfnamefont {S.}~\bibnamefont
  {Ikeda}}, \bibinfo {author} {\bibfnamefont {J.}~\bibnamefont {Hayakawa}},
  \bibinfo {author} {\bibfnamefont {Y.~M.}\ \bibnamefont {Lee}}, \bibinfo
  {author} {\bibfnamefont {F.}~\bibnamefont {Matsukura}}, \bibinfo {author}
  {\bibfnamefont {Y.}~\bibnamefont {Ohno}}, \bibinfo {author} {\bibfnamefont
  {T.}~\bibnamefont {Hanyu}}, \ and\ \bibinfo {author} {\bibfnamefont
  {H.}~\bibnamefont {Ohno}},\ }\href@noop {} {\bibfield  {journal} {\bibinfo
  {journal} {IEEE Transactions on Electron Devices}\ }\textbf {\bibinfo
  {volume} {54}},\ \bibinfo {pages} {991} (\bibinfo {year} {2007})}\BibitemShut
  {NoStop}%
\bibitem [{\citenamefont {Zhu}\ and\ \citenamefont {Park}(2006)}]{Park2006}%
  \BibitemOpen
  \bibfield  {author} {\bibinfo {author} {\bibfnamefont {J.-G.~J.}\
  \bibnamefont {Zhu}}\ and\ \bibinfo {author} {\bibfnamefont {C.}~\bibnamefont
  {Park}},\ }\href@noop {} {\bibfield  {journal} {\bibinfo  {journal}
  {Materials today}\ }\textbf {\bibinfo {volume} {9}},\ \bibinfo {pages} {36}
  (\bibinfo {year} {2006})}\BibitemShut {NoStop}%
\bibitem [{\citenamefont {Lenz}\ and\ \citenamefont
  {Edelstein}(2006)}]{Edelstein2006}%
  \BibitemOpen
  \bibfield  {author} {\bibinfo {author} {\bibfnamefont {J.}~\bibnamefont
  {Lenz}}\ and\ \bibinfo {author} {\bibfnamefont {A.~S.}\ \bibnamefont
  {Edelstein}},\ }\href@noop {} {\bibfield  {journal} {\bibinfo  {journal}
  {IEEE Sensors Journal}\ }\textbf {\bibinfo {volume} {6}},\ \bibinfo {pages}
  {631} (\bibinfo {year} {2006})}\BibitemShut {NoStop}%
\bibitem [{\citenamefont {Katayama}\ \emph {et~al.}(2006)\citenamefont
  {Katayama}, \citenamefont {Yuasa}, \citenamefont {Velev}, \citenamefont
  {Zhuravlev}, \citenamefont {Jaswal},\ and\ \citenamefont
  {Tsymbal}}]{Tsymbal2006}%
  \BibitemOpen
  \bibfield  {author} {\bibinfo {author} {\bibfnamefont {T.}~\bibnamefont
  {Katayama}}, \bibinfo {author} {\bibfnamefont {S.}~\bibnamefont {Yuasa}},
  \bibinfo {author} {\bibfnamefont {J.}~\bibnamefont {Velev}}, \bibinfo
  {author} {\bibfnamefont {M.~Y.}\ \bibnamefont {Zhuravlev}}, \bibinfo {author}
  {\bibfnamefont {S.~S.}\ \bibnamefont {Jaswal}}, \ and\ \bibinfo {author}
  {\bibfnamefont {E.~Y.}\ \bibnamefont {Tsymbal}},\ }\href@noop {} {\bibfield
  {journal} {\bibinfo  {journal} {Appl. Phys. Lett.}\ }\textbf {\bibinfo
  {volume} {89}},\ \bibinfo {pages} {112503} (\bibinfo {year}
  {2006})}\BibitemShut {NoStop}%
\bibitem [{\citenamefont {van~der Heijden}\ \emph {et~al.}(1997)\citenamefont
  {van~der Heijden}, \citenamefont {Bloemen}, \citenamefont {Metselaar},
  \citenamefont {Wolf}, \citenamefont {Gaines}, \citenamefont {van Eemeren},
  \citenamefont {van~der Zaag},\ and\ \citenamefont {de~Jonge}}]{Jonge1997}%
  \BibitemOpen
  \bibfield  {author} {\bibinfo {author} {\bibfnamefont {P.~A.~A.}\
  \bibnamefont {van~der Heijden}}, \bibinfo {author} {\bibfnamefont {P.~J.~H.}\
  \bibnamefont {Bloemen}}, \bibinfo {author} {\bibfnamefont {J.~M.}\
  \bibnamefont {Metselaar}}, \bibinfo {author} {\bibfnamefont {R.~M.}\
  \bibnamefont {Wolf}}, \bibinfo {author} {\bibfnamefont {J.~M.}\ \bibnamefont
  {Gaines}}, \bibinfo {author} {\bibfnamefont {J.~T. W.~M.}\ \bibnamefont {van
  Eemeren}}, \bibinfo {author} {\bibfnamefont {P.~J.}\ \bibnamefont {van~der
  Zaag}}, \ and\ \bibinfo {author} {\bibfnamefont {W.~J.~M.}\ \bibnamefont
  {de~Jonge}},\ }\href@noop {} {\bibfield  {journal} {\bibinfo  {journal}
  {Phys. Rev. B}\ }\textbf {\bibinfo {volume} {55}},\ \bibinfo {pages} {11569}
  (\bibinfo {year} {1997})}\BibitemShut {NoStop}%
\bibitem [{\citenamefont {Faure-Vincent}\ \emph {et~al.}(2002)\citenamefont
  {Faure-Vincent}, \citenamefont {Tiusan}, \citenamefont {Bellouard},
  \citenamefont {Popova}, \citenamefont {Hehn}, \citenamefont {Montaigne},\
  and\ \citenamefont {Schuhl}}]{Schuhl2002}%
  \BibitemOpen
  \bibfield  {author} {\bibinfo {author} {\bibfnamefont {J.}~\bibnamefont
  {Faure-Vincent}}, \bibinfo {author} {\bibfnamefont {C.}~\bibnamefont
  {Tiusan}}, \bibinfo {author} {\bibfnamefont {C.}~\bibnamefont {Bellouard}},
  \bibinfo {author} {\bibfnamefont {E.}~\bibnamefont {Popova}}, \bibinfo
  {author} {\bibfnamefont {M.}~\bibnamefont {Hehn}}, \bibinfo {author}
  {\bibfnamefont {F.}~\bibnamefont {Montaigne}}, \ and\ \bibinfo {author}
  {\bibfnamefont {A.}~\bibnamefont {Schuhl}},\ }\href@noop {} {\bibfield
  {journal} {\bibinfo  {journal} {Phys. Rev. B}\ }\textbf {\bibinfo {volume}
  {89}},\ \bibinfo {pages} {107206} (\bibinfo {year} {2002})}\BibitemShut
  {NoStop}%
\bibitem [{\citenamefont {Popova}\ \emph {et~al.}(2007)\citenamefont {Popova},
  \citenamefont {Keller}, \citenamefont {Gendron}, \citenamefont {Tiusan},
  \citenamefont {Schuhl},\ and\ \citenamefont {Lesnik}}]{Lesnik2007}%
  \BibitemOpen
  \bibfield  {author} {\bibinfo {author} {\bibfnamefont {E.}~\bibnamefont
  {Popova}}, \bibinfo {author} {\bibfnamefont {N.}~\bibnamefont {Keller}},
  \bibinfo {author} {\bibfnamefont {F.}~\bibnamefont {Gendron}}, \bibinfo
  {author} {\bibfnamefont {C.}~\bibnamefont {Tiusan}}, \bibinfo {author}
  {\bibfnamefont {A.}~\bibnamefont {Schuhl}}, \ and\ \bibinfo {author}
  {\bibfnamefont {N.~A.}\ \bibnamefont {Lesnik}},\ }\href@noop {} {\bibfield
  {journal} {\bibinfo  {journal} {Appl. Phys. Lett.}\ }\textbf {\bibinfo
  {volume} {91}},\ \bibinfo {pages} {112504} (\bibinfo {year}
  {2007})}\BibitemShut {NoStop}%
\bibitem [{\citenamefont {Heinonen}, \citenamefont {Stokes},\ and\
  \citenamefont {Yi}(2010)}]{Yi2010}%
  \BibitemOpen
  \bibfield  {author} {\bibinfo {author} {\bibfnamefont {O.~G.}\ \bibnamefont
  {Heinonen}}, \bibinfo {author} {\bibfnamefont {S.}~\bibnamefont {Stokes}}, \
  and\ \bibinfo {author} {\bibfnamefont {J.}~\bibnamefont {Yi}},\ }\href@noop
  {} {\bibfield  {journal} {\bibinfo  {journal} {Phys. Rev. Lett.}\ }\textbf
  {\bibinfo {volume} {105}},\ \bibinfo {pages} {066602} (\bibinfo {year}
  {2010})}\BibitemShut {NoStop}%
\bibitem [{\citenamefont {Hammerling}\ \emph {et~al.}(2003)\citenamefont
  {Hammerling}, \citenamefont {Zabloudil}, \citenamefont {Weinberger},
  \citenamefont {Lindner}, \citenamefont {Kosubek}, \citenamefont {Nunthel},\
  and\ \citenamefont {Baberschke}}]{Baberschke2003}%
  \BibitemOpen
  \bibfield  {author} {\bibinfo {author} {\bibfnamefont {R.}~\bibnamefont
  {Hammerling}}, \bibinfo {author} {\bibfnamefont {J.}~\bibnamefont
  {Zabloudil}}, \bibinfo {author} {\bibfnamefont {P.}~\bibnamefont
  {Weinberger}}, \bibinfo {author} {\bibfnamefont {J.}~\bibnamefont {Lindner}},
  \bibinfo {author} {\bibfnamefont {E.}~\bibnamefont {Kosubek}}, \bibinfo
  {author} {\bibfnamefont {R.}~\bibnamefont {Nunthel}}, \ and\ \bibinfo
  {author} {\bibfnamefont {K.}~\bibnamefont {Baberschke}},\ }\href@noop {}
  {\bibfield  {journal} {\bibinfo  {journal} {Phys. Rev. B}\ }\textbf {\bibinfo
  {volume} {68}},\ \bibinfo {pages} {092406} (\bibinfo {year}
  {2003})}\BibitemShut {NoStop}%
\bibitem [{\citenamefont {Moritz}\ \emph {et~al.}(2004)\citenamefont {Moritz},
  \citenamefont {Garcia}, \citenamefont {Toussaint}, \citenamefont {Dieny},\
  and\ \citenamefont {Nozieres}}]{Nozieres2004}%
  \BibitemOpen
  \bibfield  {author} {\bibinfo {author} {\bibfnamefont {J.}~\bibnamefont
  {Moritz}}, \bibinfo {author} {\bibfnamefont {F.}~\bibnamefont {Garcia}},
  \bibinfo {author} {\bibfnamefont {J.~C.}\ \bibnamefont {Toussaint}}, \bibinfo
  {author} {\bibfnamefont {B.}~\bibnamefont {Dieny}}, \ and\ \bibinfo {author}
  {\bibfnamefont {J.~P.}\ \bibnamefont {Nozieres}},\ }\href@noop {} {\bibfield
  {journal} {\bibinfo  {journal} {Europhys. Lett.}\ }\textbf {\bibinfo {volume}
  {65}},\ \bibinfo {pages} {123} (\bibinfo {year} {2004})}\BibitemShut
  {NoStop}%
\bibitem [{\citenamefont {Schrag}\ \emph {et~al.}(2000)\citenamefont {Schrag},
  \citenamefont {Anguelouch}, \citenamefont {Ingvarsson}, \citenamefont {Xiao},
  \citenamefont {Lu}, \citenamefont {Trouilloud}, \citenamefont {Gupta},
  \citenamefont {Wanner}, \citenamefont {Gallagher}, \citenamefont {Rice},\
  and\ \citenamefont {Parkin}}]{Parkin2000}%
  \BibitemOpen
  \bibfield  {author} {\bibinfo {author} {\bibfnamefont {B.~D.}\ \bibnamefont
  {Schrag}}, \bibinfo {author} {\bibfnamefont {A.}~\bibnamefont {Anguelouch}},
  \bibinfo {author} {\bibfnamefont {S.}~\bibnamefont {Ingvarsson}}, \bibinfo
  {author} {\bibfnamefont {G.}~\bibnamefont {Xiao}}, \bibinfo {author}
  {\bibfnamefont {Y.}~\bibnamefont {Lu}}, \bibinfo {author} {\bibfnamefont
  {P.~L.}\ \bibnamefont {Trouilloud}}, \bibinfo {author} {\bibfnamefont
  {A.}~\bibnamefont {Gupta}}, \bibinfo {author} {\bibfnamefont {R.~A.}\
  \bibnamefont {Wanner}}, \bibinfo {author} {\bibfnamefont {W.~J.}\
  \bibnamefont {Gallagher}}, \bibinfo {author} {\bibfnamefont {P.~M.}\
  \bibnamefont {Rice}}, \ and\ \bibinfo {author} {\bibfnamefont {S.~S.~P.}\
  \bibnamefont {Parkin}},\ }\href@noop {} {\bibfield  {journal} {\bibinfo
  {journal} {Appl. Phys. Lett.}\ }\textbf {\bibinfo {volume} {77}},\ \bibinfo
  {pages} {2373} (\bibinfo {year} {2000})}\BibitemShut {NoStop}%
\bibitem [{\citenamefont {Kools}\ \emph {et~al.}(1999)\citenamefont {Kools},
  \citenamefont {Kula}, \citenamefont {Mauri},\ and\ \citenamefont
  {Lin}}]{Lin1999}%
  \BibitemOpen
  \bibfield  {author} {\bibinfo {author} {\bibfnamefont {J.~C.~S.}\
  \bibnamefont {Kools}}, \bibinfo {author} {\bibfnamefont {W.}~\bibnamefont
  {Kula}}, \bibinfo {author} {\bibfnamefont {D.}~\bibnamefont {Mauri}}, \ and\
  \bibinfo {author} {\bibfnamefont {T.}~\bibnamefont {Lin}},\ }\href@noop {}
  {\bibfield  {journal} {\bibinfo  {journal} {Journal of Applied Physics}\
  }\textbf {\bibinfo {volume} {85}},\ \bibinfo {pages} {4466} (\bibinfo {year}
  {1999})}\BibitemShut {NoStop}%
\bibitem [{\citenamefont {Kuznetsov}, \citenamefont {Udalov},\ and\
  \citenamefont {Fraerman}(2018)}]{Fraerman2018}%
  \BibitemOpen
  \bibfield  {author} {\bibinfo {author} {\bibfnamefont {M.~A.}\ \bibnamefont
  {Kuznetsov}}, \bibinfo {author} {\bibfnamefont {O.~G.}\ \bibnamefont
  {Udalov}}, \ and\ \bibinfo {author} {\bibfnamefont {A.~A.}\ \bibnamefont
  {Fraerman}},\ }\href@noop {} {\bibfield  {journal} {\bibinfo  {journal}
  {ArXiv: 1807.05590}\ } (\bibinfo {year} {2018})}\BibitemShut {NoStop}%
\bibitem [{\citenamefont {Tegen}\ \emph {et~al.}(2001)\citenamefont {Tegen},
  \citenamefont {Monch}, \citenamefont {Schumann}, \citenamefont {Vinzelberg},\
  and\ \citenamefont {Schneider}}]{Schneider2001}%
  \BibitemOpen
  \bibfield  {author} {\bibinfo {author} {\bibfnamefont {S.}~\bibnamefont
  {Tegen}}, \bibinfo {author} {\bibfnamefont {I.}~\bibnamefont {Monch}},
  \bibinfo {author} {\bibfnamefont {J.}~\bibnamefont {Schumann}}, \bibinfo
  {author} {\bibfnamefont {H.}~\bibnamefont {Vinzelberg}}, \ and\ \bibinfo
  {author} {\bibfnamefont {C.~M.}\ \bibnamefont {Schneider}},\ }\href@noop {}
  {\bibfield  {journal} {\bibinfo  {journal} {Journal of Applied Physics}\
  }\textbf {\bibinfo {volume} {89}},\ \bibinfo {pages} {8169} (\bibinfo {year}
  {2001})}\BibitemShut {NoStop}%
\bibitem [{\citenamefont {Egelhoff}\ \emph {et~al.}(2006)\citenamefont
  {Egelhoff}, \citenamefont {McMichael}, \citenamefont {Dennis}, \citenamefont
  {Stiles}, \citenamefont {Shapiro}, \citenamefont {Maranville},\ and\
  \citenamefont {Powell}}]{Powell2006}%
  \BibitemOpen
  \bibfield  {author} {\bibinfo {author} {\bibfnamefont {W.~F.}\ \bibnamefont
  {Egelhoff}}, \bibinfo {author} {\bibfnamefont {R.~D.}\ \bibnamefont
  {McMichael}}, \bibinfo {author} {\bibfnamefont {C.~L.}\ \bibnamefont
  {Dennis}}, \bibinfo {author} {\bibfnamefont {M.~D.}\ \bibnamefont {Stiles}},
  \bibinfo {author} {\bibfnamefont {A.~J.}\ \bibnamefont {Shapiro}}, \bibinfo
  {author} {\bibfnamefont {B.~B.}\ \bibnamefont {Maranville}}, \ and\ \bibinfo
  {author} {\bibfnamefont {C.~J.}\ \bibnamefont {Powell}},\ }\href@noop {}
  {\bibfield  {journal} {\bibinfo  {journal} {Appl. Phys. Lett.}\ }\textbf
  {\bibinfo {volume} {88}},\ \bibinfo {pages} {162508} (\bibinfo {year}
  {2006})}\BibitemShut {NoStop}%
\bibitem [{\citenamefont {Udalov}, \citenamefont {Fraerman},\ and\
  \citenamefont {Demidov}(2018)}]{Demidov2018}%
  \BibitemOpen
  \bibfield  {author} {\bibinfo {author} {\bibfnamefont {O.~G.}\ \bibnamefont
  {Udalov}}, \bibinfo {author} {\bibfnamefont {A.~A.}\ \bibnamefont
  {Fraerman}}, \ and\ \bibinfo {author} {\bibfnamefont {E.~S.}\ \bibnamefont
  {Demidov}},\ }\href@noop {} {\bibfield  {journal} {\bibinfo  {journal}
  {ArXiv: 1808.01296}\ } (\bibinfo {year} {2018})}\BibitemShut {NoStop}%
\bibitem [{\citenamefont {Topkaya}\ \emph {et~al.}(2010)\citenamefont
  {Topkaya}, \citenamefont {Erkovan}, \citenamefont {Ozturk}, \citenamefont
  {Ozturk}, \citenamefont {Aktas},\ and\ \citenamefont
  {Ozdemir}}]{Ozdemir2010}%
  \BibitemOpen
  \bibfield  {author} {\bibinfo {author} {\bibfnamefont {R.}~\bibnamefont
  {Topkaya}}, \bibinfo {author} {\bibfnamefont {M.}~\bibnamefont {Erkovan}},
  \bibinfo {author} {\bibfnamefont {A.}~\bibnamefont {Ozturk}}, \bibinfo
  {author} {\bibfnamefont {O.}~\bibnamefont {Ozturk}}, \bibinfo {author}
  {\bibfnamefont {B.}~\bibnamefont {Aktas}}, \ and\ \bibinfo {author}
  {\bibfnamefont {M.}~\bibnamefont {Ozdemir}},\ }\href@noop {} {\bibfield
  {journal} {\bibinfo  {journal} {J. Appl. Phys.}\ }\textbf {\bibinfo {volume}
  {108}},\ \bibinfo {pages} {023910} (\bibinfo {year} {2010})}\BibitemShut
  {NoStop}%
\bibitem [{\citenamefont {Liu}\ and\ \citenamefont
  {Furdyna}(2006)}]{Furdyna2006}%
  \BibitemOpen
  \bibfield  {author} {\bibinfo {author} {\bibfnamefont {X.}~\bibnamefont
  {Liu}}\ and\ \bibinfo {author} {\bibfnamefont {J.~K.}\ \bibnamefont
  {Furdyna}},\ }\href@noop {} {\bibfield  {journal} {\bibinfo  {journal} {J.
  Phys.: Condens. Matter}\ }\textbf {\bibinfo {volume} {18}},\ \bibinfo {pages}
  {R245} (\bibinfo {year} {2006})}\BibitemShut {NoStop}%
\bibitem [{\citenamefont {Nascimento}\ \emph {et~al.}(2006)\citenamefont
  {Nascimento}, \citenamefont {Saitovitch}, \citenamefont {Pelegrinia},
  \citenamefont {Figueredo}, \citenamefont {Biondo},\ and\ \citenamefont
  {Passamani}}]{Passamani2006}%
  \BibitemOpen
  \bibfield  {author} {\bibinfo {author} {\bibfnamefont {V.~P.}\ \bibnamefont
  {Nascimento}}, \bibinfo {author} {\bibfnamefont {E.~B.}\ \bibnamefont
  {Saitovitch}}, \bibinfo {author} {\bibfnamefont {F.}~\bibnamefont
  {Pelegrinia}}, \bibinfo {author} {\bibfnamefont {L.~C.}\ \bibnamefont
  {Figueredo}}, \bibinfo {author} {\bibfnamefont {A.}~\bibnamefont {Biondo}}, \
  and\ \bibinfo {author} {\bibfnamefont {E.~C.}\ \bibnamefont {Passamani}},\
  }\href@noop {} {\bibfield  {journal} {\bibinfo  {journal} {J. Appl. Phys.}\
  }\textbf {\bibinfo {volume} {99}},\ \bibinfo {pages} {08C108} (\bibinfo
  {year} {2006})}\BibitemShut {NoStop}%
\bibitem [{\citenamefont {Popova}\ \emph {et~al.}(2006)\citenamefont {Popova},
  \citenamefont {Tiusan}, \citenamefont {Schuhl}, \citenamefont {Gendron},\
  and\ \citenamefont {Lesnik}}]{Lesnik2006}%
  \BibitemOpen
  \bibfield  {author} {\bibinfo {author} {\bibfnamefont {E.}~\bibnamefont
  {Popova}}, \bibinfo {author} {\bibfnamefont {C.}~\bibnamefont {Tiusan}},
  \bibinfo {author} {\bibfnamefont {A.}~\bibnamefont {Schuhl}}, \bibinfo
  {author} {\bibfnamefont {F.}~\bibnamefont {Gendron}}, \ and\ \bibinfo
  {author} {\bibfnamefont {N.~A.}\ \bibnamefont {Lesnik}},\ }\href@noop {}
  {\bibfield  {journal} {\bibinfo  {journal} {Phys. Rev. B}\ }\textbf {\bibinfo
  {volume} {74}},\ \bibinfo {pages} {224415} (\bibinfo {year}
  {2006})}\BibitemShut {NoStop}%
\bibitem [{\citenamefont {Yang}\ \emph {et~al.}(2017)\citenamefont {Yang},
  \citenamefont {Nan}, \citenamefont {Zhang}, \citenamefont {Zhou},
  \citenamefont {Peng}, \citenamefont {Ren}, \citenamefont {Ye}, \citenamefont
  {Sun},\ and\ \citenamefont {Liu}}]{PhysRevApplied.8.044006}%
  \BibitemOpen
  \bibfield  {author} {\bibinfo {author} {\bibfnamefont {Q.}~\bibnamefont
  {Yang}}, \bibinfo {author} {\bibfnamefont {T.}~\bibnamefont {Nan}}, \bibinfo
  {author} {\bibfnamefont {Y.}~\bibnamefont {Zhang}}, \bibinfo {author}
  {\bibfnamefont {Z.}~\bibnamefont {Zhou}}, \bibinfo {author} {\bibfnamefont
  {B.}~\bibnamefont {Peng}}, \bibinfo {author} {\bibfnamefont {W.}~\bibnamefont
  {Ren}}, \bibinfo {author} {\bibfnamefont {Z.-G.}\ \bibnamefont {Ye}},
  \bibinfo {author} {\bibfnamefont {N.~X.}\ \bibnamefont {Sun}}, \ and\
  \bibinfo {author} {\bibfnamefont {M.}~\bibnamefont {Liu}},\ }\href@noop {}
  {\bibfield  {journal} {\bibinfo  {journal} {Phys. Rev. Applied}\ }\textbf
  {\bibinfo {volume} {8}},\ \bibinfo {pages} {044006} (\bibinfo {year}
  {2017})}\BibitemShut {NoStop}%
\end{thebibliography}%

\end{document}